\documentclass[a4paper,10pt,twocolumn]{article}

\usepackage{amssymb}
\usepackage{amsmath}
\usepackage[figuresright]{rotating}
\usepackage{multirow}

\usepackage{booktabs}
\rmfamily
\begin{document}
\begin{titlepage}
%\begin{frontmatter}




%\title{Alice results from LHC\\* Heavy-flavor production in LHC pp interactions using the ALICE detector}
\title{Heavy-flavor production in LHC pp interactions using the ALICE detector}


%\author{Bj$\o$rn S. Nilsen\\* On behalf of the ALICE Collaboration}
%\address{Department of Physics, Creighton University, Omaha, NE 68178 U.S.A.}
\author{Bj\o rn S. Nilsen\\* On behalf of the ALICE Collaboration\\Department of Physics, Creighton University, Omaha, NE 68178 U.S.A.}

\maketitle

\rule{6in}{0.5mm}
\begin{abstract}
Measurements of charm and beauty production in pp collisions, using the ALICE
detector system, at LHC energies ($\sqrt{s} = 2.76$ and $7.0$ TeV) can test perturbative
QCD down to very low  Bj\"{o}rken-x. They are also critical as a reference to
ALICE's heavy ion program. The ALICE detector system allows measurements
not covered by the other LHC experiments in addition to covering
complementary regions. A description of the ALICE detector system, in relation
to ATLAS and CMS,  are presented. Results from both leptonic and hadronic
decay channels will be shown along with comparisons to other measurements
when available.

\vspace{0.25in}
\begin{flushleft}
\end{flushleft}
\end{abstract}
\rule{6in}{0.5mm}

%\begin{keyword}
%Hadron-Hardron Scattering \sep
%LHC \sep
%ALICE experiment \sep
%ultra-relativistic heavy ion collisions \sep
%heavy flavour production \sep
%nuclear modification factor \sep
%pp collisions \sep
%single muons \sep
%single electrons


%\end{keyword}

%\end{frontmatter}
\end{titlepage}
%\linenumbers

\section{Introduction}
\label{1}
The Quark Gluon Plasma (QGP) produced by colliding lead ions together using
the Large Hadron Collider (LHC) will get hot enough to produce, from the
QCD vacuum, up, down, and quite likely strange quarks \cite{thermalstrange}. At LHC energies, 
all quarks can be produced in the initial interactions between the 
lead ions, specifically the heavier charm, bottom, and top quarks. The
production of hadrons containing these quarks can be measured in 
proton-proton (pp) interactions at equivalent energies. When such 
quarks or hadrons, created in lead-lead interactions, pass though a 
QGP their energy and hadronic properties are very likely to be modified 
by the QGP in ways that can tell us a lot about the properties of the 
QGP phase of matter.

The presentations of Donald Isenhower and Michael Murray at
this conference gave a very good introduction and review of QGP
formation and the role of heavy flavor 
measurements, also see \cite{ALICE-LHC}. For a proper understanding
of the results from heavy flavor measurements in lead-lead 
interactions, good
measurements of charm production in pp collisions are necessary, preferably with
the same instruments having the same acceptance and other systematics. 
In addition the production cross sections and decay rates of heavy flavored 
hadrons are of interest on their own. Here we present the most recent 
results from the A Large Ion Collider Experiment (ALICE) detector on 
open charm measurements in pp collisions at both $\sqrt{s}=7$ 
and $2.76$ TeV.
\section{ALICE Detector}
\label{2}
The ALICE detector has been designed specifically to study the 
properties of the QGP. To this end, it has very good momentum 
resolution at very low transverse momentum, as low at 80 MeV/c,
and very good particle identification. All of this is achieved 
by taking advantage of the 
varied capabilities of the 19 different sub-detector types, but
at the cost of a limited acceptance and data taking rate. This is in
contrast to the other LHC experiments, which are optimized for other
measurements. This results in ALICE contributing measurements uniquely in
the low momentum region with a significantly better particle identification
capability. In this way the measurements of ATLAS and CMS, both 
in Pb-Pb and 
pp interactions, are supplemented by those of ALICE into regions not
reachable by ATLAS or CMS.

\begin{figure}[htbp] %  figure placement: here, top, bottom, or page
   \centering
   \includegraphics[width=3in,clip=true,trim=0in 0in 0in 0in]{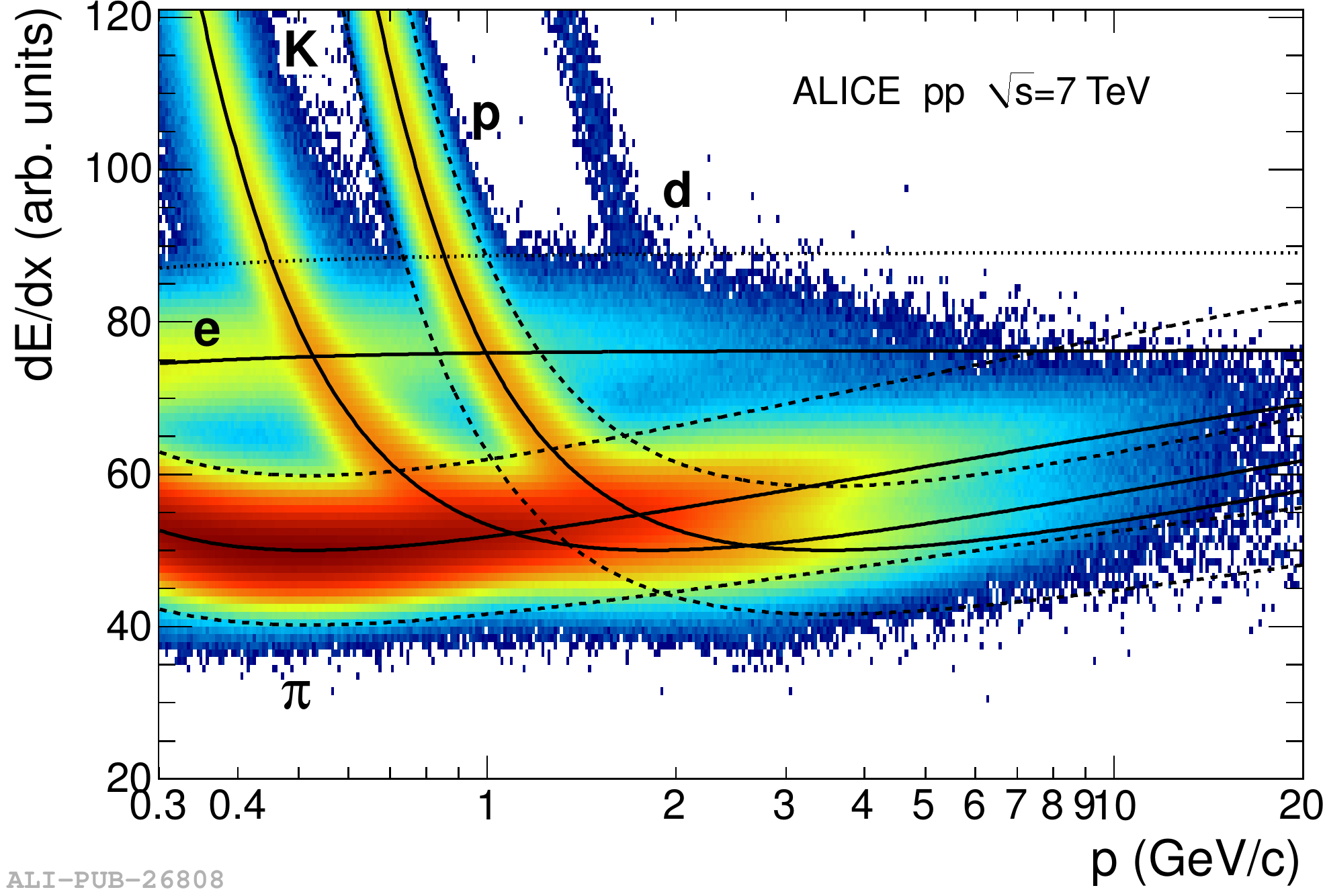} 
   \caption{Specific energy loss in the TPC as a function of momentum 
   with superimposed Bethe-Bloch lines for various particle species. 
   The dashed lines show the pion and proton exclusion bands. The 
   dotted line corresponds to the $+3\sigma$ cut for electrons (see 
   \cite{ALICE-JPsi-pp7}).}
   \label{fig:tpcdedxPID}
\end{figure}

\begin{figure}[!t] %  figure placement: here, top, bottom, or page
   \centering
   \includegraphics[width=3in,clip=true,trim=0in 3.2in 4in 0.25in]{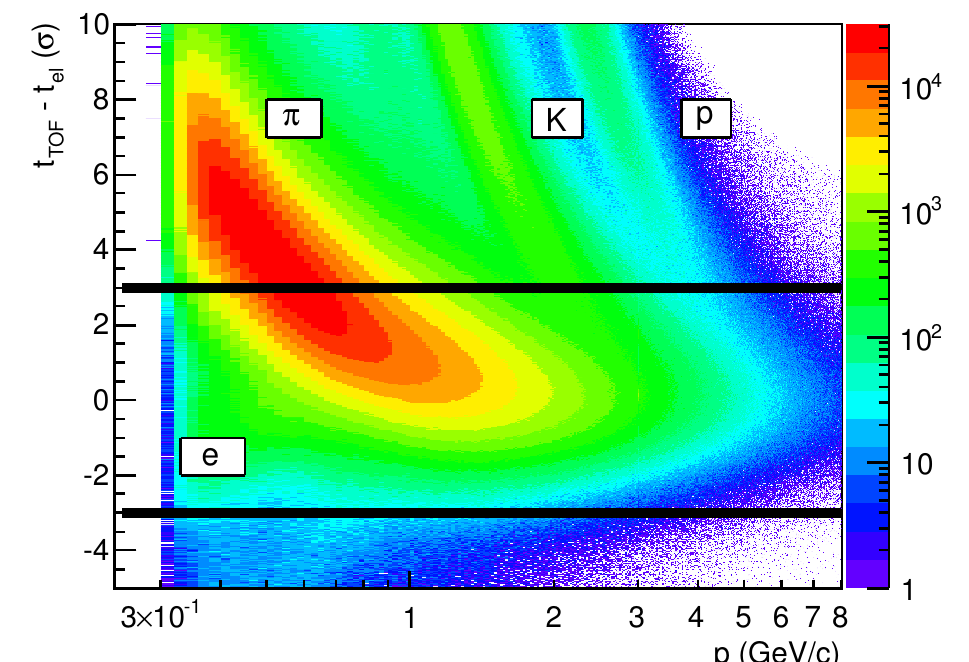}
   \includegraphics[width=3in,clip=true,trim=4in 3.2in 0in 0.25in]{Figures/2012-May-26-eID_TOFTRDTPC-eps-converted-to.pdf} 
   \caption{Electron identification for the TPC-TRD-TOF analysis.
   Upper shows a pid plot based on TOF-T0 PID.
   The lower shows the electron hadron separation from the TRD.
   %The lower right show the TRD cut used.
   %Lower shows the resulting TPC ionization signal vs. track momentum 
   %after Kaon, Proton/anti-proton, and deuteron tracks removed. 
   See \cite{ALICE-HF-eDecaypp7}  for more details.}
   \label{fig:tpctoftrdPIDA}
\end{figure}

\begin{figure}[!t]
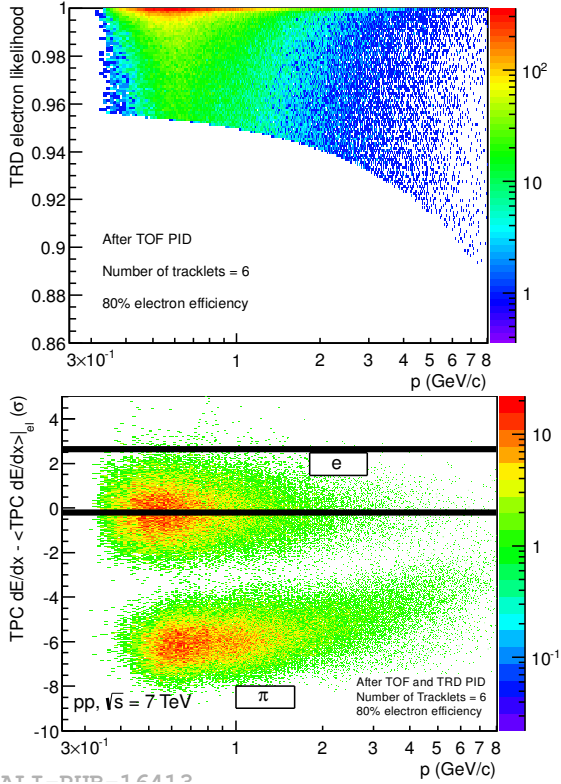
 %  figure placement: here, top, bottom, or page
   \centering
   \includegraphics[width=3in,clip=true,trim=4in 0.1in 0in 3.2in]{Figures/2012-May-26-eID_TOFTRDTPC-eps-converted-to.pdf} 
   \includegraphics[width=3in,clip=true,trim=0in 0.1in 4in 3.4in]{Figures/2012-May-26-eID_TOFTRDTPC-eps-converted-to.pdf}
   \caption{Electron identification for the TPC-TRD-TOF analysis.
   %Upper left shows a pid plot based on TOF-T0 PID.
   %The upper shows the electron hadron separation from the TRD.
   The upper show the TRD cut used.
   Lower left shows the resulting TPC ionization signal vs. track 
   momentum after Kaon, Proton/anti- proton, and deuteron tracks removed.
   See \cite{ALICE-HF-eDecaypp7}  for more details.}
   \label{fig:tpctoftrdPIDB}
\end{figure}

In the analysis presented here, measurements from the sub-detectors listed 
in Table \ref{tab:sub-Dets} where used extensively. The V0 and SPD 
sub-detectors are primarily used for triggering, along with the 
$\mu-$Trigger chambers. For non-$\mu$ events, those
taken without the $\mu-$Tracker, the central barrel Time Projection 
Chamber (TPC) was the primary track, momentum, and particle identification
detector. Figure \ref{fig:tpcdedxPID} shows the correlation between the ionization of the
TPC gas associated with a track and the track's momentum.
In addition parts of the Inner Tracking System (ITS) where used both 
for triggering and particle identification. The ITS is made up of
2 inner cylindrical layers of silicon pixel detectors (SPD), which can also be use
for triggering, 2 middle cylindrical layers of silicon drift detectors (SDD), and the
2 outer most cylindrical layers of silicon double sided micro-strip detectors (SSD).
Other detectors, such as the Time
of Flight (TOF), Transition Radiation Detector (TRD), or even the
Electromagnetic Calorimeter (EMCal) where used, primarily for
electron/positron identification. See Figures
\ref{fig:tpctoftrdPIDA} \& \ref{fig:tpctoftrdPIDB} where signals from the TOF and TRD are shown
and the result after removing kaons, protons/ant-protons, and deuterons 
is seen on the TPC ionization vs momentum correlation plot.

Muons are measured in the region $-4.0<\eta<-2.5$. In front of
these tracking chambers is a 10 interaction length hadron absorber
which lets through muons above about 4 GeV/c. Such muons
will trigger the muon trigger chambers which are behind an additional
7.2 interaction length absorber. Between these two absorbers are a
series of muon tracking chambers imbedded in a 3 Tm magnetic field to measure
these muon's momentum. Further details of the ALICE detectors can be found 
in \cite{ALICE-LHC}.

\section{Acceptance and Data selection}
\label{3}
This paper presents pp data at $\sqrt{s}=$ 7 and 2.76 TeV taken during 2010
and 2011 respectively. The physical acceptance of the ALICE detectors
(relevant to this analysis) is shown in Table \ref{tab:sub-Dets}. During
the shutdown between the 2010 and 2011 data taking periods, additional
TRD and EMCal super modules were installed.

Data analyzed was taken from minimum bias and $\mu$-minimum bias runs. The minimum bias
trigger required at least 1 signal in the SPD detectors ($|\eta|<1.95$) or one
signal in either V0 sub-detector and these signals were required to occur during
a beam-beam crossing. The $\mu$-minimum bias runs have the same requirement with the
addition of a signal in the $\mu$-trigger detector. Van der Meer
scans were used to measure the total minimum bias cross section \cite{vandermeerscan}.
These cross sections were found to be $62.5\pm2.2$ mb at 7 TeV and
$54.8\pm1.7$ mb at 2.76 TeV. The luminosity depends a bit on the specific data selected
for analysis and are shown in Figures \ref{fig:DmassPlots}, \ref{fig:DcrossSectionPlots},
and \ref{fig:CharmToTheory7TeV}.

\begin{table}[!b]
   \begin{tabular}{@{}l@{}c@{$\,$}c@{$\!\!$}r@{}} % Column formatting, @{} suppresses leading/trailing space
      \hline\hline
         Sub- &  \multicolumn{2}{c}{Acceptance} & \\ \cline{2-3}
      Detector & $\eta$                & $\phi$ & Use \\ \hline
      V$0_{A}$ & $+2.8\rightarrow+5.1$ & $2\pi$ & \multirow{2}{*}{Trgr $t_{0}$} \\
      V$0_{C}$ & $-3.7\rightarrow-1.7$ & $2\pi$ & \\
      SPD      & $-2.0\rightarrow+2.0$ & $2\pi$ & Trgr, Vtx \\
      T$0_{A}$ & $+4.5\rightarrow+5.0$ & $2\pi$ & \multirow{2}{*}{TOF $t_{0}$} \\
      T$0_{C}$ & $-3.3\rightarrow-2.9$ & $2\pi$ & \\
      TPC      & $-0.9\rightarrow+1.9$ & $2\pi$ & trk, $p_{t}$, PID \\
      TOF      & $-0.9\rightarrow+1.9$ & $2\pi$ & PID \\
      \multirow{2}{*}{TRD}& \multirow{2}{*}{$-0.9\rightarrow+1.9$} & $-\frac{\pi}{9}\rightarrow\frac{2\pi}{9}$\&
      $\frac{7\pi}{9}\rightarrow\frac{11\pi}{9}^{\dagger}$ &\multirow{2}{*}{$e$ ID}\\
               &  & $-\frac{3\pi}{9}\rightarrow\frac{2\pi}{9}$\&$\frac{7\pi}{9}\rightarrow\frac{4\pi}{3}^{\ddagger}$  & \\
      \multirow{2}{*}{EMCal}&\multirow{2}{*}{$-0.7\rightarrow+0.7$} &$\frac{\pi}{2}\rightarrow\frac{11\pi}{18}^{\dagger}$ & \multirow{2}{*}{$e$/hadron ID} \\
             &                  &$\frac{\pi}{2}\rightarrow\frac{8\pi}{9}^{\ddagger}$  & \\
      $\mu$-Tkr & $-4.0\rightarrow-2.5$ & $2\pi$ & $p_{t}$ \& trk \\
      $\mu$-Tgr & $-4.0\rightarrow-2.5$ & $2\pi$ & Trgr \\ \hline
      \multicolumn{4}{l}{$^{\dagger}$ During 2010 Data Taking.} \\
      \multicolumn{4}{l}{$^{\ddagger}$ During 2011 Data Taking.} \\ \hline\hline
   \end{tabular}
   \caption{The acceptance and ALICE sub-detectors used in
   this analysis. Trgr: Trigger Detector, Vtx: Vertex finding detector,
   trk: Tracking detector, $p_{T}$ Transverse Momentum measuring detector,
   $T0$: Start time detector for TOF measurement, e ID: electron ID 
   detector, e/hadron ID: electron from hadron identification detector,
   and PID: Particle identification detector.}
   \label{tab:sub-Dets}
\end{table}

For analysis in the central barrel, tracks were required to have at least 70
out of 159 measurements in the TPC with a $\chi^{2}/{\rm ndf}<2$, and at least one
good measurement in the ITS SPD of the ITS. In
addition, when TOF timings were used for the PID, the best timings require
a signal in one of the T0 sub-detectors (or beam-beam crossing time are
tried if no T0 signals).

Acceptance calculations were done using PYTHIA 6.4.21 \cite{pythia} with the
Perugia-0 tune \cite{pythiaPerugia-0}. Events were then passed through a full ALICE simulation
matched to the run conditions (AliRoot). AliRoot used the Geant-3 
\cite{geant3} particle transport engine.
Where necessary, feed down from B meson decay has been corrected based on a Fixed Order 
Next to Leading Log (FONLL) \cite{FONLL} calculation.

\section{Results}
\label{4}
\subsection{D meson Measurements}
\label{4.1}
\subsubsection{Production cross sections}
\label{4.1.1}

\begin{figure}[!b] %  figure placement: here, top, bottom, or page
   \centering
   \includegraphics[width=3in]{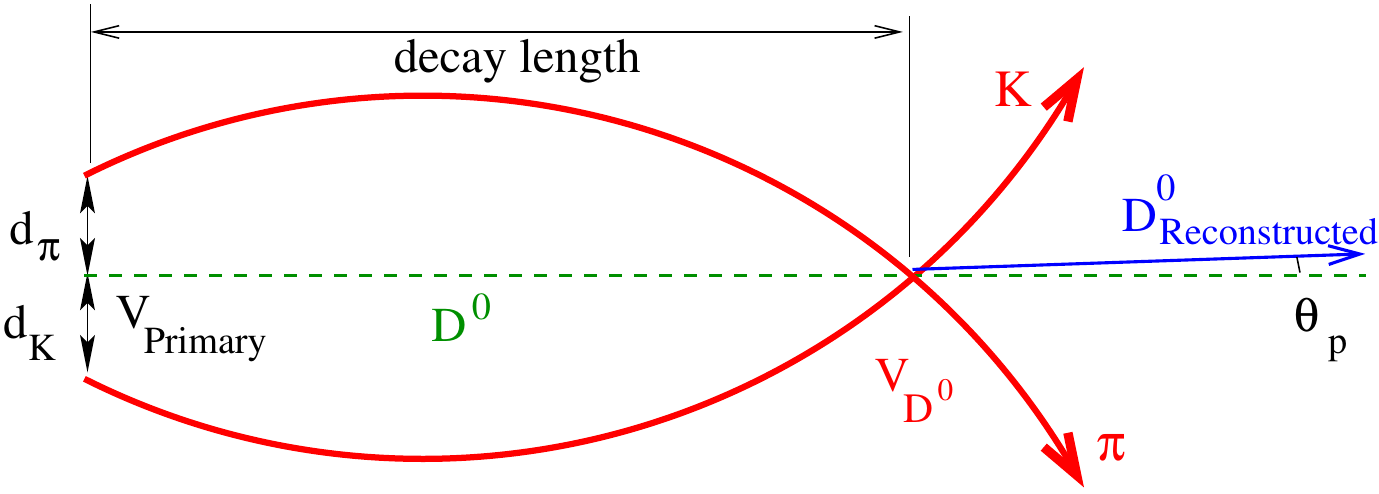} 
   \caption{Decay topology for ${\rm D^{0}}$ decay. Other D meson decay have
   similar topologies, but with a different number of final states.}
   \label{fig:DmesonDecay}
\end{figure}

The following decay have been reconstructed, ${\rm D^{0}}\rightarrow{\rm K}^{\pm}\pi^{\mp}$,
${\rm D}^{\pm}\rightarrow{\rm K}^{\mp}2\pi^{\pm}$, ${\rm D}^{\star\pm}\rightarrow {\rm D^{0}}\pi^{\pm}
\rightarrow ({\rm K}^{\pm}\pi^{\mp} \| {\rm K}^{\mp}\pi^{\pm})\pi^{\pm}$,
and ${\rm D}^{\pm}_{\rm s}\rightarrow \phi \pi^{\pm}\rightarrow {\rm K}^{\pm}{\rm K}^{\mp}\pi^{\pm}$. Note
decays of ${\rm D}^{\pm}_{\rm s}\rightarrow {\rm K}^{\star{\rm 0}}{\rm K}^{\pm}\rightarrow ({\rm K}^{\pm}\pi^{\mp}\|
{\rm K}^{\mp}\pi^{\pm}){\rm K}^{\pm}$ can also pass these selection but at a rate smaller by a factor
of about 100 (due to acceptance, efficiency, and branching ratios). 
Given the relatively long lifetimes for these D meson 
decays (${\rm D^{0}}:c\tau=122.9\mu{\rm m}$ ${\rm D}^{\pm}:311.8\mu{\rm m}$ and ${\rm D}^{\pm}_{s}:149.9\mu{\rm m}$) 
and the very good vertex resolution of ALICE, a selection based on the reconstructed
displaced vertex was used to significantly improve the signal to background ratio
in these analyses, see Figure \ref{fig:DmesonDecay}. Particle Identification was
based on signals in the TPC, and TOF, see Figures 
\ref{fig:tpcdedxPID}, \ref{fig:tpctoftrdPIDA} and \ref{fig:tpctoftrdPIDB} above.

\begin{figure}[htbp] %  figure placement: here, top, bottom, or page
   \centering
   \includegraphics[width=3in,clip=true,trim=0in 0in 2.4in 0.4in]{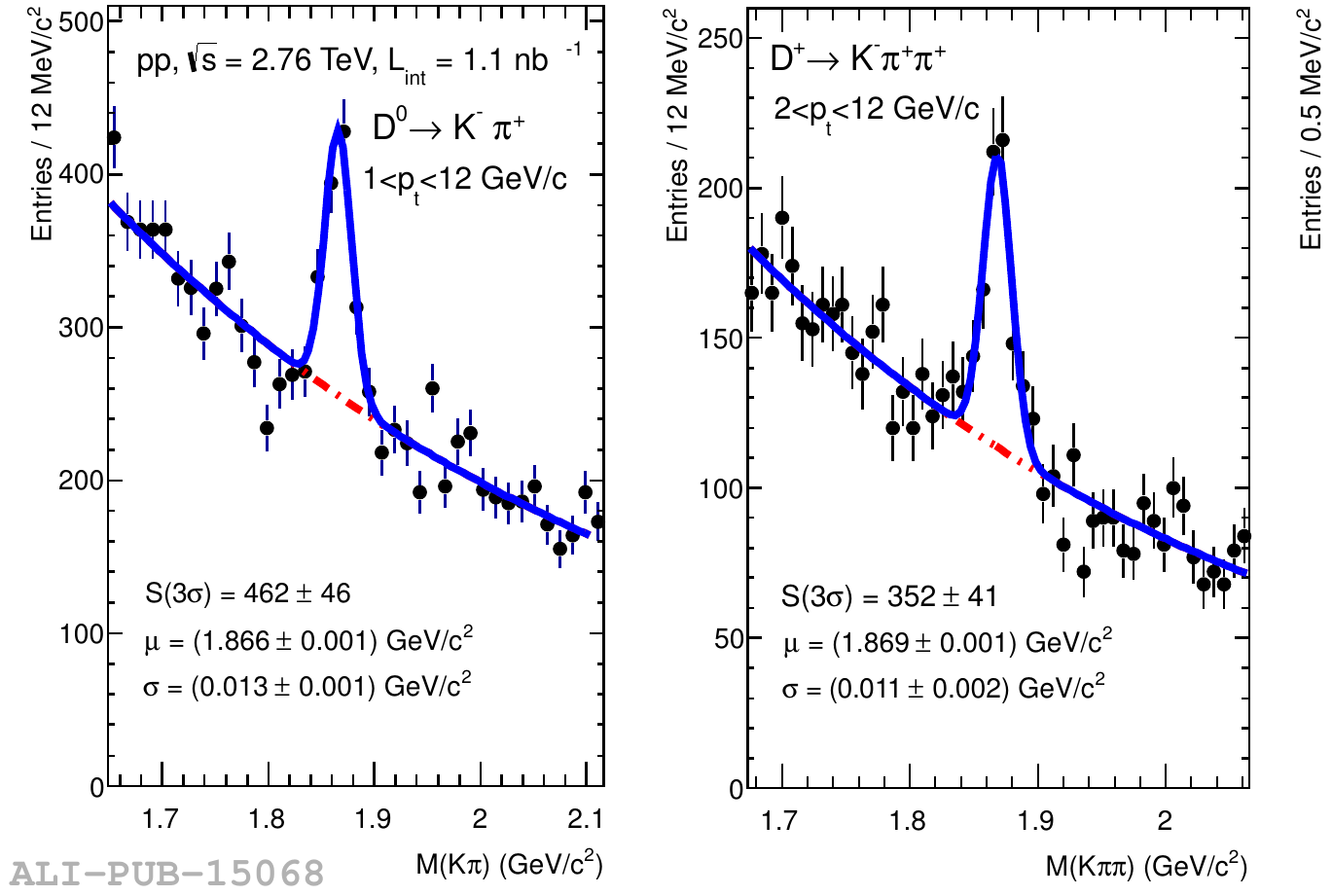}
   \includegraphics[width=1.5in,clip=true,trim=5.3in 0.05in 0in 0.4in]{Figures/2012-Jul-26-Dmesons276_massPlots-eps-converted-to.pdf} 
   \caption{Invariant-mass spectrum of ${\rm D^{0}} + \bar{\rm D^{0}}$ (upper left) 
   and ${\rm D}^{+} + {\rm D}^{-}$ (upper right) 
    candidates, and invariant-mass difference, 
    $\Delta m=m({\rm K}\pi\pi)-m({\rm K}\pi)$, for 
    ${\rm D}^{\star+} + {\rm D}^{\star-}$ candidates (lower) in pp collisions 
    at $\sqrt{s} = 2.76\;$TeV (see \cite{ALICE-Charm-pp2.76}).}
   \label{fig:DmassPlots}
\end{figure}

\begin{figure}[htbp] %  figure placement: here, top, bottom, or page
   \centering
   \includegraphics[width=3in]{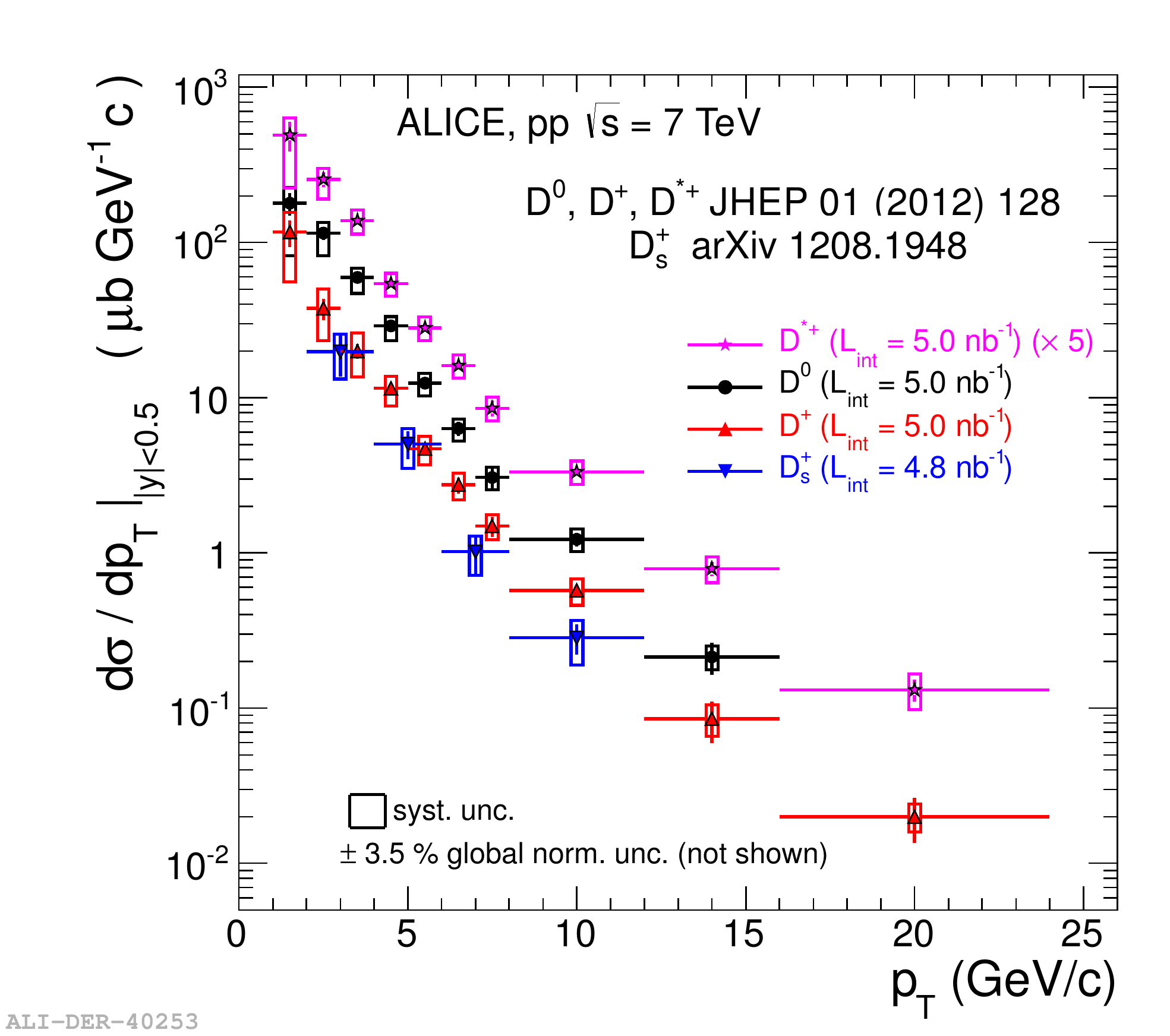} 
   \caption{$p_{T}$-differential cross section of ${\rm D^{0}}$,${\rm D}^{+}$, 
   ${\rm D}^{\star +}$ and ${\rm D}^{+}_{\rm s}$ measured with the 2010 pp collisions 
   at 7 TeV.}
   \label{fig:DcrossSectionPlots}
\end{figure}

Mass plots were then produced from the candidate decay products, for the
${\rm D}^{\star\pm}$ a mass difference between the 3 decay products ($M({\rm K}\pi\pi)
-M({\rm K}\pi)$) was plotted. The background distributions were fit using a polynomial up to
second degree, and a Gaussian distribution for the signals. The yield was then measured
as the area in the Gaussian distribution. An example is shown in Figure \ref{fig:DmassPlots}
from the 2.76 TeV pp data set. The $p_{T}$ dependent cross at 7 TeV is shown
in Figure \ref{fig:DcrossSectionPlots}. A detailed comparison to FONLL and the
General Mass Variable Flavor Number Schema (GM-VFNS) \cite{GMVFNS} calculation has been 
done, not shown, and the agreement between these
results and both theories are within their theoretical uncertainties \cite{ALICE-Charm-pp7,
ALICE-Charm-pp2.76,ALICE-Ds-pp7}.

\subsubsection{D meson production ratios}
\label{4.1.2}
Ratios of particle production rates have the advantage that many systematic
uncertainties cancel out and can give some insight to differences in
production mechanisms. In addition some particle interactions simulations,
like PYTHIA \cite{pythia}, use such ratios as inputs to their production
simulations. One such ratio is the ratio between spin 1 mesons to spin 0
mesons for charm and heaver mesons (PYTHIA 6.4.21's PARJ(13)). Given
the measurements above we have a measure of the fraction $P_{v}$ of ${\rm c}\bar{\rm d}$ D meson production 
in a vector to those produced in a vector plus pseudoscalar state. We get
\[\begin{array}{ll}P_{v}(7\;{\rm TeV\; pp})=&\!\!\!\!0.59\pm 0.06({\rm st})\pm 0.08({\rm sy})\\
&\!\!\!\! \pm 0.010({\rm BR})^{+0.005}_{-0.003}({\rm ex}) \end{array} \]
and
\[\begin{array}{ll}P_{v}(2.76\;{\rm TeV\; pp})=&\!\!\!\!0.65\pm 0.10({\rm st})\pm 0.08({\rm sy}) \\
&\!\!\!\! \pm 0.010({\rm BR})^{+0.011}_{-0.004}({\rm ex})\end{array}\]
(uncertainties are from st=statistical, sy=systimatic, BR=branching ratio, 
and ex=due to extrapolation to full phase space)\cite{ALICE-Charm-pp2.76}. 
These values compare well with those measured
by ATLAS at 7 TeV \cite{ATLAS_Pv}, CDF at 1.96 TeV \cite{CDF_Pv}, 
ALEPH at 91.2 GeV \cite{ALEPH_Pv}, and CLEO at 10.55 GeV \cite{CLEO_Pv}.

\subsection{Leptonic decays of D and B mesons}
\label{4.2}
\begin{figure}[htbp] %  figure placement: here, top, bottom, or page
   \centering
   \includegraphics[width=3in,trim=0in 0.1in 0in 0.1in,clip=true]{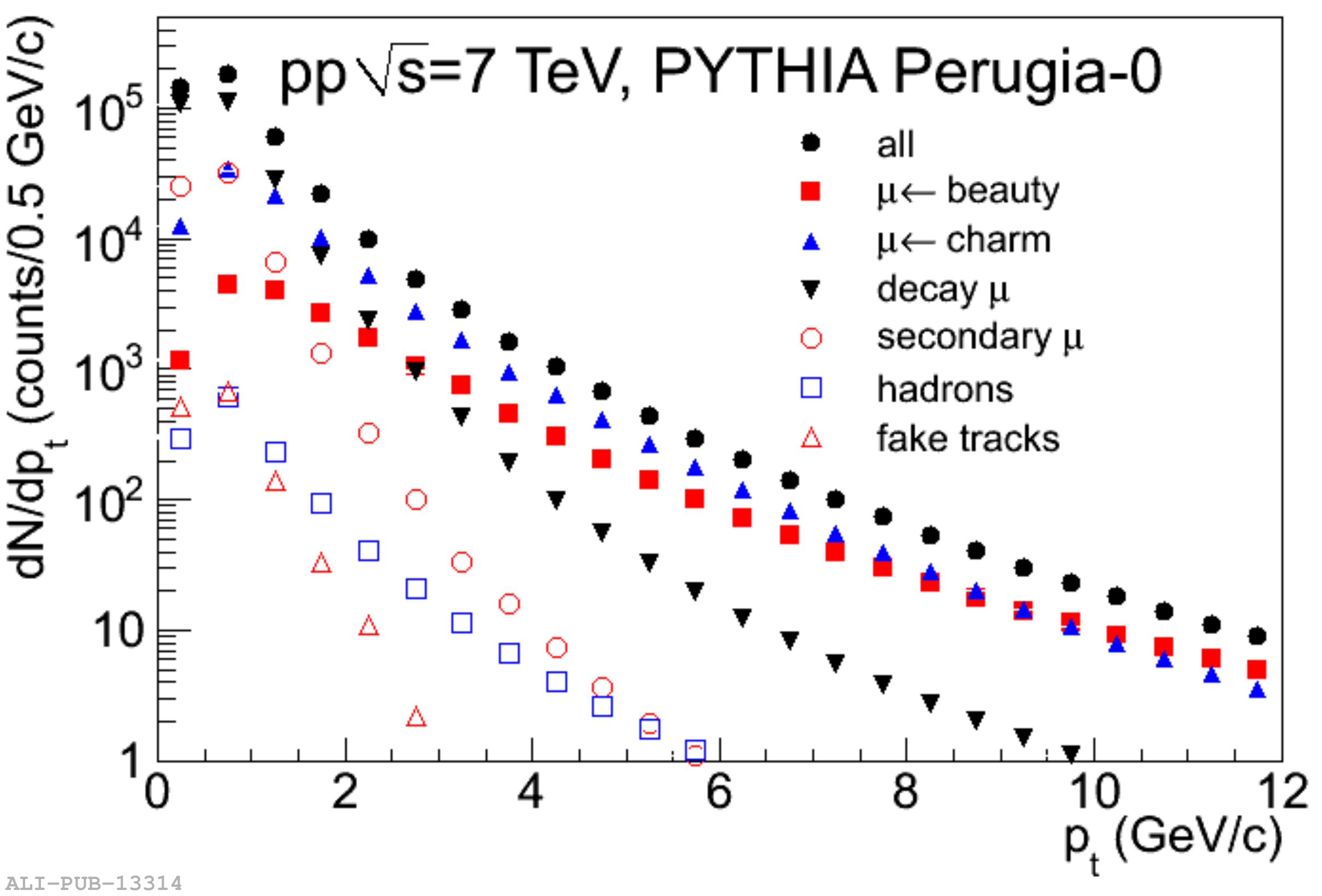} 
   \caption{Transverse momentum distribution of reconstructed 
   tracks in the muon spectrometer after all selection cuts 
   were applied. The distributions were obtained from a PYTHIA 
   \cite{pythia} (tune Perugia-0 \cite{pythiaPerugia-0}) simulation 
   of pp collisions at $\sqrt s$ = 7 TeV. The main sources are 
   indicated in the figure.
   (see \cite{ALICE-HF-MuDecaypp7}).}
   \label{fig:MuonSources}
\end{figure}

All charm and bottom quark mesons have decays which include
leptons in their final states. These leptons can be rather
energetic and tend to dominate all other sources of leptons,
but to get a more accurate measure of the D and B mesons a
proper cocktail from all of the sources of leptons needs to
be considered. For final state muons, this is shown in Figure
\ref{fig:MuonSources}. A somewhat more complicated cocktail has
been done for sources of electrons (which can also be created
by interactions with elements of the ALICE detector).

\subsubsection{$\mu$ from D and B decays}
\label{4.2.1}
\begin{figure}[!t] %  figure placement: here, top, bottom, or page
   \centering
   \includegraphics[width=2.8in,trim=0in 0in 9.46in 0.1in,clip=true]{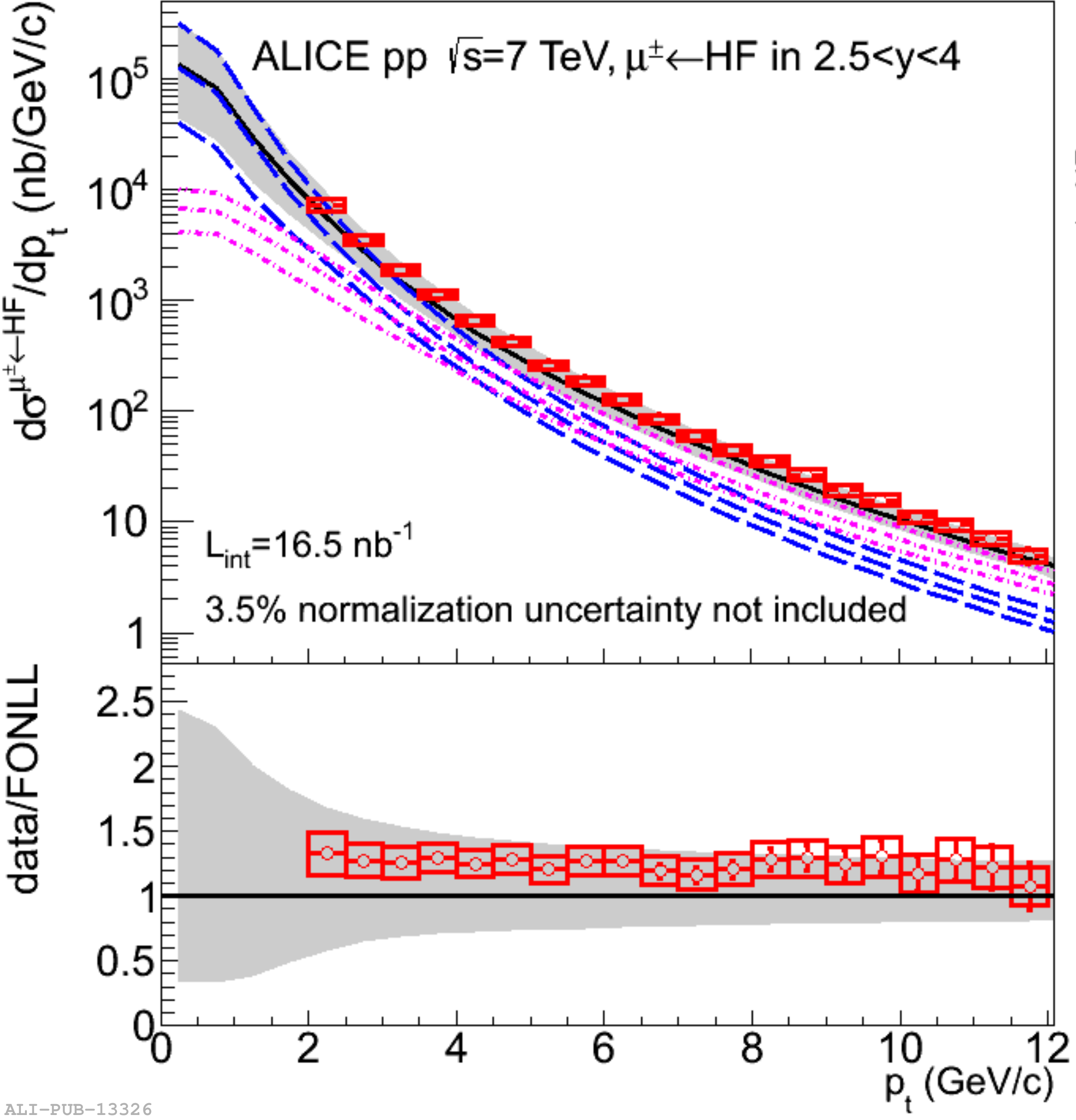} 
   \includegraphics[width=2.9in,trim=9.41in 0.14in -0.2in 0in,clip=true]{Figures/2012-Feb-08-fig3-eps-converted-to.pdf} 
   \caption{$p_{T}$-differential (top) and y-differential (bottom) 
   production cross section of muons from heavy flavour decays. 
   In both panels, the error bars (empty boxes) represent the 
   statistical (systematic) uncertainties. The solid curves are 
   FONLL calculations and the bands display the theoretical 
   systematic uncertainties. Also shown, are the FONLL calculations 
   \cite{FONLL} and systematic theoretical uncertainties for 
   muons from charm (long dashed curves) and beauty (dashed curves) 
   decays. The lower panels show the corresponding ratios between 
   data and FONLL calculations.
   (see \cite{ALICE-HF-MuDecaypp7}).}
   \label{fig:CharmToTheory7TeV}
\end{figure}

The $p_{T}$ and $y$ dependent cross sections for muons directly from
charm and beauty decay, produced in 7 TeV pp interactions, are shown in 
Figure \ref{fig:CharmToTheory7TeV}. A simular plot for 2.76 TeV pp
interactions can be found in \cite{ALICE-HF-MuDecayPbPbpp2.76}. In
both cases the measurements are compared to FONLL calculations
and are found to be within theoretical uncertainties.

\subsubsection{${\rm e}$ from D and B decays}
\label{4.2.2}
\begin{figure}[htbp] %  figure placement: here, top, bottom, or page
   \centering
   \includegraphics[width=3in,clip=true,trim=0in 0.05in 0in 0.65in]{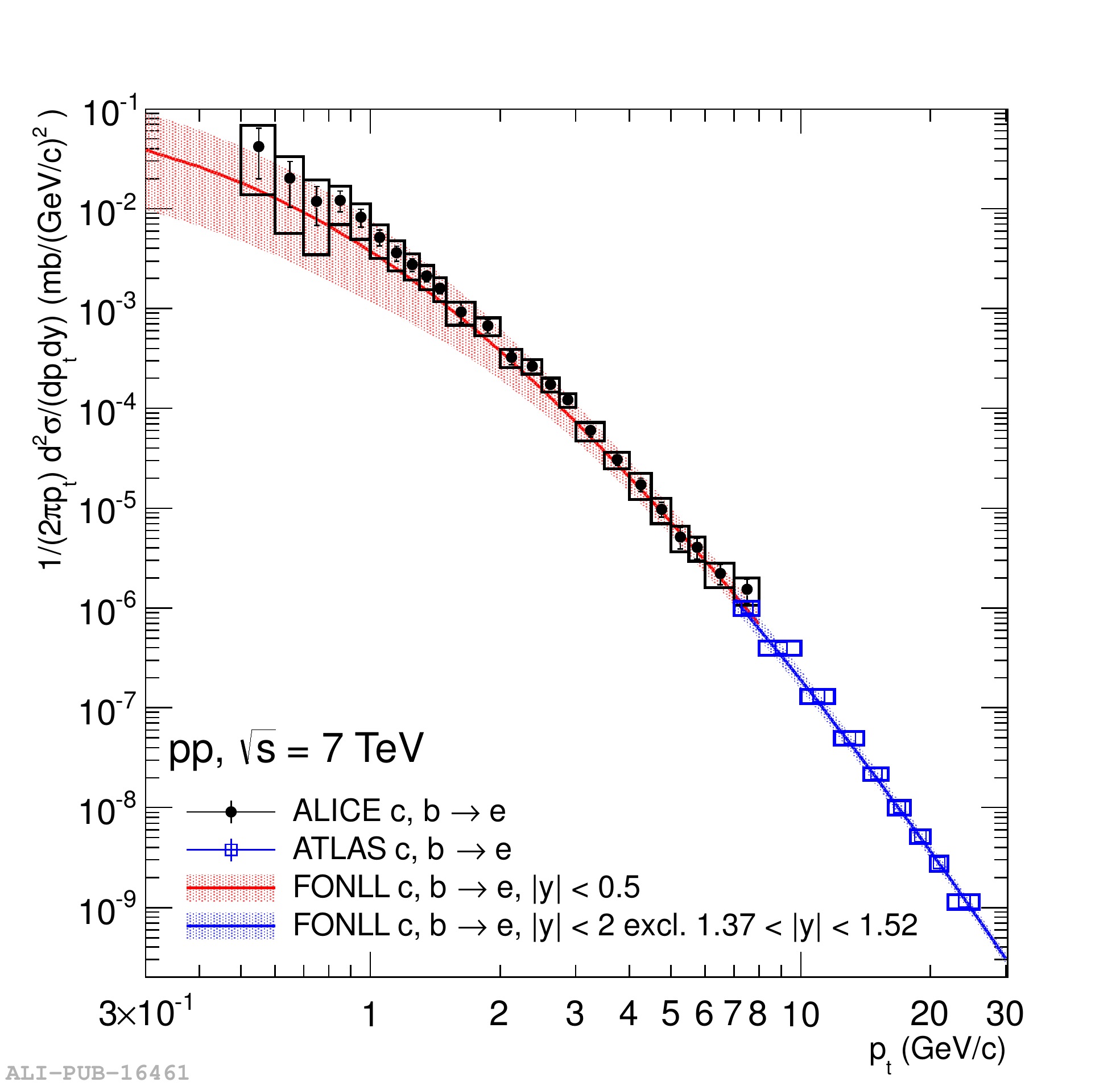} 
   \caption{Invariant differential production cross sections of electrons from heavy-flavour decays
   measured by ALICE and ATLAS \cite{ATLAS-eandmu7tev} in pp collisions at
   $\sqrt{s}=7\;TeV$ in different rapidity intervals. FONL pQCD calculations 
   with the same rapidity selections are shown for comparison.
   (see \cite{ALICE-HF-eDecaypp7}).}
   \label{fig:CharmToe7TeV}
\end{figure}

Electrons/positrons decay from charm and beauty have also been
measured for 7 TeV pp collisions, see \cite{ALICE-HF-eDecaypp7}.
These results also are well described by a FONLL calculation, 
Figure \ref{fig:CharmToe7TeV}. Also shown
in this figure are similar measurements from ATLAS \cite{ATLAS-eandmu7tev}.
The ATLAS results extend the ALICE results very nicely.

\subsubsection{Charm production cross section}
\label{4.2.3}
\begin{figure}[htbp] %  figure placement: here, top, bottom, or page
   \centering
   \includegraphics[width=3in,clip=true,trim=0in 0in 0in 0.3in]{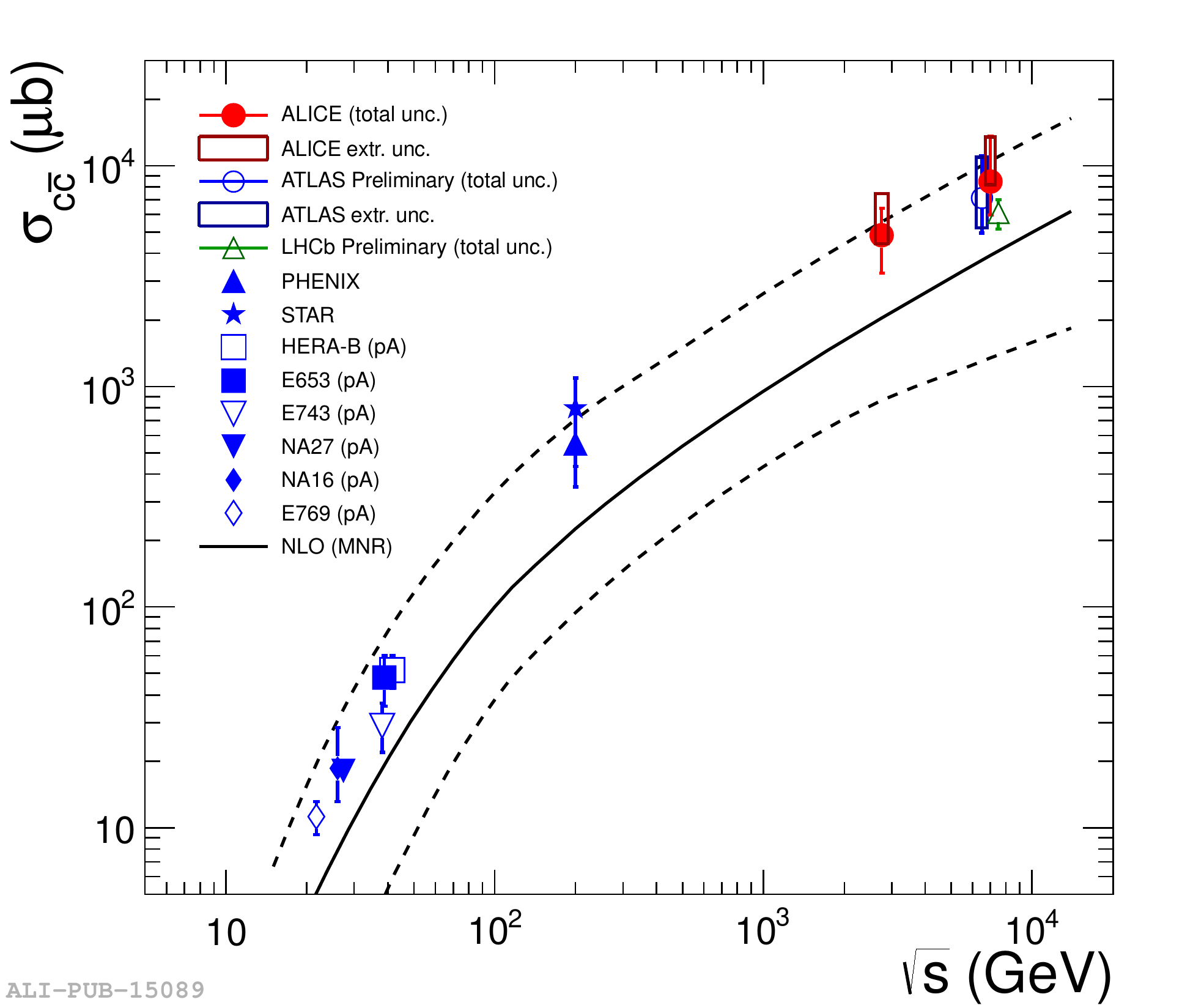} 
   \caption{Energy dependence of the total nucleonÐ-nucleon charm 
   production cross section. In case of protonÐ-nucleus (p--A) or 
   deuteronÐ-nucleus (dA) collisions, the measured cross sections 
   have been scaled down by the number of binary nucleonÐ-nucleon 
   collisions calculated in a Glauber model of the protonÐ-nucleus 
   or deuteronÐ-nucleus collision geometry. The NLO MNR calculation 
   (and its uncertainties) \cite{mangano-NLO MNR} is represented by solid (dashed) lines.
   (see \cite{ALICE-Charm-pp2.76}).}
   \label{fig:CharmEdep}
\end{figure}

The total charm production cross section for 7 and 2.75 TeV pp
collisions are shown in Figure \ref{fig:CharmEdep}
\[\begin{array}{ll}
\sigma^{tot}_{\rm c\bar{c}}(7\;{\rm TeV})=&\!\!\!\!8.5\pm 0.5({\rm st})^{+1.0}_{-2.4}({\rm sy})\pm 0.1({\rm BR})\\
&\!\!\!\!\pm 0.2({\rm FF})\pm 0.3({\rm lum})^{+5.0}_{-0.4}({\rm ex})\; {\rm mb}\end{array}\]
and
\[\begin{array}{ll}\sigma^{tot}_{\rm c\bar{c}}(2.76\;{\rm TeV})=&\!\!\!\!4.8\pm 0.8({\rm st})^{+1.0}_{-1.3}({\rm sy})\pm 0.06({\rm BR})\\
&\!\!\!\!\pm 0.1({\rm FF})\pm 0.1({\rm lum})^{+2.6}_{-0.4}({\rm ex})\; {\rm mb} \end{array}\]
along with other measurements (uncertainties are from FF=the fragmentation fraction $\equiv$
relative production yield for a charm quark hadronizing to a particular species of D meson, 
and lum=the luminosity). Proton nucleus and deuteron nucleus data are
also shown but scaled down by the number of binary collisions in
these reactions, as computed by a Glauber model. An extrapolation
to the full phase space has been done. The solid line
is a NLO MNR calculation (uncertainties indicated by the dashed 
lines) \cite{mangano-NLO MNR}. Note that all of the measurements from the LHC experiments
agree within their uncertainties and those at other energies are a bit above but
seem to follow the energy dependence from the NLO MNR calculation.

\section{Conclusions}
\label{5}
ALICE has measured up to 4 D-meson production cross sections at up to 2 different
values of $\sqrt{s}$ in pp collisions near central rapidity. We have also measured the heavy flavor
production in these pp collisions, centrally via electron/positron decay channels
and more forwardly via muon decay channels. By extrapolating these
measurements to the full phase space, we have obtained the total 
charm production cross. All of these measurements are
described within uncertainties by pQCD calculations.

%% References

\nocite{*}
\bibliographystyle{elsarticle-num}
\bibliography{martin}



\end{document}